\documentstyle[12pt,amssymb]{article}

\begin{document}

\tolerance=5000

\def\pp{{\, \mid \hskip -1.5mm =}}
\def\cL{{\cal L}}
\def\be{\begin{equation}}
\def\ee{\end{equation}}
\def\bea{\begin{eqnarray}}
\def\eea{\end{eqnarray}}
\def\beq{\begin{eqnarray}}
\def\eeq{\end{eqnarray}}
\def\tr{{\rm tr}\, }
\def\nn{\nonumber \\}
\def\e{{\rm e}}

\begin{titlepage}

\ 

\vfill 

\begin{center}
\Large
{\bf Multisupergravity from latticized extra dimension
}

\vfill

\normalsize

Shin'ichi Nojiri$^\spadesuit$\footnote{Electronic mail: nojiri@nda.ac.jp, 
snojiri@yukawa.kyoto-u.ac.jp},
Sergei D. Odintsov$^{\heartsuit\clubsuit}$\footnote{Electronic mail: 
odintsov@ieec.fcr.es Also at TSPU, Tomsk, Russia}

\

\normalsize

\vfill

{\em $\spadesuit$ Department of Applied Physics, 
National Defence Academy, \\
Hashirimizu Yokosuka 239-8686, JAPAN}

\ 

{\em $\heartsuit$ Institut d'Estudis Espacials de Catalunya (IEEC), \\
Edifici Nexus, Gran Capit\`a 2-4, 08034 Barcelona, SPAIN}

\ 

{\em $\clubsuit$ Instituci\`o Catalana de Recerca i Estudis 
Avan\c{c}ats (ICREA), \\Barcelona, SPAIN}

\end{center}

\vfill 

\baselineskip=24pt

\begin{abstract}

The construction of the linearized  four-dimensional multisupergravity from
five-dimensional linearized supergravity with discretized fifth dimension
is presented. The one-loop vacuum energy is evaluated when (anti)periodic 
boundary conditions are chosen for (bosons) fermions, respectively or
vice-versa. It is proposed that the relation between discretized M-theory 
and strings may be found in the same fashion.

\end{abstract}



\end{titlepage}

\section{Introduction}

It is known that the earlier attempts to construct  multigravity theories \cite{ISS} 
were not quite successful. The problem is that for massive  tensor theories 
the consistency issue is not yet completely understood. 
Such theories normally contain several patologies like van Dam-Veltman-Zakharov
discontinuity \cite{DVZ} or the appearence of ghosts beoynd quadratic order.
The truncation of KK theory to finite number of massive spin-2 fields is not
consistent too \cite{duff}.
 
Recently, the interesting approach to multigravity theories was initiated 
by brane-world picture. Such approach suggests a new way to formulate
multigravity where it is possible to check how the consistency of 
brane-world gravity is kept at discretized level.
In ref.\cite{AGS}, related with the deconstruction of extra dimension, a model 
where several four-dimensional gravities are connected by the link variables
has been proposed. 
By choosing a proper gauge condition, it has been shown that the mass term of the 
graviton is generated. 
Almost in parallel, in \cite{KS}, starting from the linearized Einstein theory in 
five dimensions and replacing the fifth direction with a lattice, a four-dimensional 
model with massive multi-graviton has been proposed. This has been generalized 
by introducing the non-nearest neighbour couplings on the lattice \cite{CENOZ}. 
In the model\cite{KS}, the one-loop vacuum energy is negative in general but in the 
model of \cite{CENOZ}, the vacuum energy can be positive. This may explain 
the acceleration of the present universe. The consistency of discretized gravity 
(up to quadratic order), its continuum limit and some solutions have been studied 
in recent works \cite{DM}.
The attractive property of multigravity is related with the fact that 
it may lead to interesting (still acceptable!) modifications of Newton law
at large scales.

In this paper, we (super)generalize the models\cite{CENOZ}, where theory 
\cite{KS} is also included,  starting from the linearized supergravity in 
five dimensions. Of course, the latticized action should respect most of 
the symmetries of the five-dimensional action.
As usually, there is the hope that consistency issue appearing in bosonic 
version may be resolved within more fundamental, supersymmetric theory. 
The resulting multisupergravity which includes multigravitons and multigravitinos
can be obtained by replacing the fifth direction in the linearized supergravity 
to the discrete lattice. 

\section{Multisupergravity from extra dimension}

Before deconstructing supergravity, we consider the five-dimensional linearized 
supergravity. In the five dimensions, there is no Majorana representation of the 
$\gamma$-matrices and the Rarita-Schwinger field (gravitino) should be complex. 
Therefore, reducing the five-dimensional supergravity \cite{5dSG} to the
four-dimensional one, the obtained supergravity has at least ${\cal N}=2$ 
(local) supersymmetry. 
Furhtermore, as there is no chiral fermion in the five dimensions, there is 
no chiral multiplet. The minimum representation of the multiplet including the 
graviton is associated with the complex Rarita-Shwinger field and $U(1)$ vector 
(gauge) field. The number of bosonic degrees of  freedom is
$5$(graviton)$+3$(vector)$=8$. 
On the other hand, the number of fermionic degrees of freedom coming from complex 
Rarita-Schwinger field is $4\times 2=8$ (2 comes from the complexity). 

The action of the five dimensional linearized supergravity is given by
considering the perturbation from the flat background which is the 
vacuum. The explicit form is 
\bea
\label{dSG1}
S&=&\int d^5 x\left\{ {\cal L}_E + {\cal L}_{RS} + {\cal L}_{U} \right\}\ ,\\
\label{dSG2}
{\cal L}_E&=&-{1 \over 2}\partial_\lambda h_{\mu\nu}\partial^\lambda h^{\mu\nu}
+ \partial_\lambda h^\lambda_{\ \mu}\partial_\nu h^{\mu\nu} 
 - \partial_\mu h_{\mu\nu}\partial_\nu h + {1 \over 2}\partial_\lambda h
\partial^\lambda h\ ,\\
\label{dSG3}
{\cal L}_{RS} &=& i\bar \psi_\mu \gamma^{\mu\nu\rho}\partial_\nu \psi_\rho\ , \\
\label{dSG4} 
{\cal L}_U &=& - {1 \over 4}F^{\mu\nu}F_{\mu\nu}\ .
\eea
Here $h_{\mu\nu}$ is the graviton and $h\equiv h^{\ \mu}_\mu$. The Rarita-Schwinger 
field is denoted by $\psi_\mu$. The $U(1)$ field strength $F_{\mu\nu}$ is given by 
the gauge (vector) field $A_\mu$ as usual: $F_{\mu\nu}=\partial_\mu A_\nu 
 - \partial_\nu A_\mu$. We may introduce the fluctuation ${\cal E}_\mu^{\ a}$ of the 
f\" unfbein field $e_\mu^{\ a}$ from the flat background: $e_\mu^{\ a}=\delta_\mu^{\ a}
+ {\cal E}_\mu^{\ a}$. In terms of ${\cal E}_\mu^{\ a}$, one may rewrite $h_{\mu\nu}$ 
as $h_{\mu\nu}=\eta_{\mu a}{\cal E}_\nu^{\ a} + {\cal E}_\mu^{\ a}\eta_{a \mu}$ 
(we may choose the gravitational coupling constant $\kappa$ to be unity here). 
Here $\eta_{\mu a}=\delta_\mu^{\ b}\eta_{ba}$ and 
$\eta_{a \mu}=\eta_{ab}\delta_\mu^{\ b}$. The metric tensor $\eta_{ab}$ 
($a,b=0,1,2,3,4$) in the flat local Lorentz space is given by
\be
\label{dSG5}
\eta_{ab}=\left(\begin{array}{ccccc}
-1 & 0 & 0 & 0 & 0 \\
0 & 1 & 0 & 0 & 0 \\
0 & 0 & 1 & 0 & 0 \\
0 & 0 & 0 & 1 & 0 \\
0 & 0 & 0 & 0 & 1 
\end{array}\right)\ .
\ee
As the action (\ref{dSG2}) is written in terms of $h_{\mu\nu}$, if one
expresses $h_{\mu\nu}$ with the help of ${\cal E}_\mu^{\ a}$, the action 
is invariant under the local Lorentz transformation 
\be
\label{dSG5b}
\delta {\cal E}_\mu^{\ a}=\omega^{ab}\eta_{b\mu}\ ,
\ee
where $\omega^{ab}$ is local parameter with $\omega^{ab}=-\omega^{ba}$. Under the 
transformation, $h_{\mu\nu}$ is invariant. By the transformation (\ref{dSG5b}), 
the gauge condition may be chosen 
\be
\label{dSG5c}
{1 \over 2}h_{\mu\nu}=\eta_{\mu a}{\cal E}_\nu^{\ a} = {\cal E}_\mu^{\ a}\eta_{a \mu}\ .
\ee
In the linearized gravity, by using the gauge condition (\ref{dSG5c}), we may forget 
${\cal E}_\mu^{\ a}$. 

The action (\ref{dSG1}) is invariant under the linearized general coordinate 
transformation:
\be
\label{dSG6}
\delta h_{\mu\nu}=\partial_\mu \epsilon_\nu + \partial_\nu \epsilon_\mu\ ,\quad 
\delta\psi_\mu = \delta A_\mu =0\ ,
\ee
local supersymmetry transformation:
\be
\label{dSG7}
\delta h_{\mu\nu}=\delta A_\mu = 0\ ,\quad \delta \psi_\mu = \partial_\mu \eta\ ,
\ee
and $U(1)$ gauge transformation
\be
\label{dSG8}
\delta h_{\mu\nu}=\delta \psi_\mu = 0\ ,\quad A_\mu = \partial_\mu \sigma\ .
\ee
Here $\epsilon_\mu$, $\eta$, and $\sigma$ are the local parameters for the 
transformations. 
Even for the local supersymmetry transformation, the transformations do not 
mix the different kinds of fields. This is because the local transformations 
correspond to the inhomogeneous part of the full transformaions. 
We should note, however, that there is a {\it global} supersymmetry, which
mixes the fields:
\bea
\label{dSG9}
\delta h_{\mu\nu}&=& {1 \over 2}\left( \bar\zeta \gamma_\mu \psi_\nu 
+ \bar\zeta \gamma_\nu \psi_\mu  - \bar \psi_\mu \gamma_\nu \zeta 
 - \bar \psi_\nu \gamma_\mu \zeta\right)\ ,\nn
\delta \psi_\mu &=& - {i \over 4}\gamma^{\rho\sigma}\omega_{\mu\rho\sigma}\zeta 
+ F_{\rho\sigma}\left(\gamma^{\rho\sigma}\gamma_\mu - 5 \gamma^\rho \delta_\mu^{\ \sigma} 
+ 5 \gamma^\sigma \delta_\mu^{\ \rho}\right)\zeta \ ,\nn
\delta A_\mu &=& 80 \left(\bar\zeta \psi_\mu + \bar \psi_\mu \zeta\right)\ .
\eea
Here $\zeta$ is a constant spinor which is the parameter of the transformation and 
the spin connection $\omega_{\mu\nu\rho}$ is defined by
\be
\label{dSG9b}
\omega_{\mu\nu\rho}\equiv {1 \over 2}\left\{-e_{\rho a}\left(\partial_\nu e^{\ a}_\mu
 - \partial_\mu e^{\ a}_\nu\right) - e_{\mu a}\left(\partial_\rho e^{\ a}_\nu 
 - \partial_\nu e^{\ a}_\rho\right) + e_{\nu a}\left(\partial_\mu e^{| a}_\rho 
 - \partial_\rho e^{\ a}_\mu \right)\right\} \ ,
\ee
and linearized in (\ref{dSG9}):
\be
\label{dSG9c}
\omega_{\mu\nu\rho}\sim {1 \over 2} \left\{\partial_\mu h_{\rho\nu} - \partial_\rho h_{\mu\nu} 
 - \eta_{\rho a}\partial_\nu {\cal E}^{\ a}_\mu 
+ \eta_{\mu a}\partial_\nu {\cal E}^{\ a}_\rho \right\}\ .
\ee
Furthermore, with the gauge condition (\ref{dSG5c}) one gets
\be
\label{dSG9d}
\omega_{\mu\nu\rho}\sim {1 \over 2} \left\{\partial_\mu h_{\rho\nu} - \partial_\rho h_{\mu\nu} 
\right\}\ .
\ee

We now consider the deconstruction by replacing fifth spacelike dimension by 
discrete $N$ points, which may be regarded as the one-dimensional lattice. 
There were many works on realization the supersymmetry on the lattice
\cite{lSUSY,SNsusy} (for recent progress, see \cite{lSUSYrecent}). 
The interesting idea to put the supersymmetry on the finite lattice has been 
developed in \cite{SNsusy}. The problem comes from the difficulty to realize 
the Leibniz rule on the lattice. Under the supersymmetry transformation, 
the variation of the Lagrangian density becomes a total derivative by summing up 
the variation of the field by the Leibniz rule, then such action is invariant. 
However, the Leibniz rule does not hold for the difference operator on the lattice 
in general. Let us consider one-dimensional lattice and denote the difference 
operator by $\Delta$. In general,
\be
\label{dSG10}
\sum_n \left\{\left(\Delta \phi^{(1)}_n\right) \phi^{(2)}_n \phi^{(3)}_n 
+ \phi^{(1)}_n \left(\Delta \phi^{(2)}_n\right) \phi^{(3)}_n 
+ \phi^{(1)}_n \phi^{(2)}_n \left(\Delta \phi^{(3)}_n\right)\right\}\neq 0\ .
\ee
Here the point (site) on the lattice is denoted by $n$ and $\phi^{(1,2,3)}$'s 
are variables (fields) defined on the sites. We should note, however, that it is 
not difficult to realize the supersymmetry on the lattice for the free theory, 
since we only require the anti-Hermiticity for $\Delta$: 
\be
\label{dSG11}
\sum_n \left\{\left(\Delta \phi^{(1)}_n\right) \phi^{(2)}_n 
+ \phi^{(1)}_n \left(\Delta \phi^{(2)}_n\right) \right\}= 0\ .
\ee
A set satisfying (\ref{dSG11}) has been given in \cite{SNsusy}. 
We consider $N$ variables, $\phi_n$, which may be identified with the fields on a lattice 
with $N$ sites. The difference operator $\Delta$ is defined by
\be
\label{KS1}
\Delta \phi_n \equiv \sum_{k=0}^{N-1}a_k \phi_{n+k}\ .
\ee
Here it is assumed $\phi_{n+N}=\phi_n$, which may be regarded as a 
periodic boundary condition. Since
\be
\label{KS1b}
\sum_{n=0}^{N-1}\phi^{(1)}\Delta \phi_n^{(2)} = \sum_{n, k=0}^{N-1}\phi^{(1)}_n
a_k \phi_{n+k}^{(2)} = \sum_{n,k=0}^{N-1} a_{-k}\phi^{(1)}_{n+k} \phi^{(2)}_n\ ,
\ee
if 
\be
\label{KS1c}
a_{-k}\left(=a_{N-k}\right)=-a_k\ ,
\ee
Eq.(\ref{dSG11}) can be satisfied. Note that there is no nontrivial solution 
in (\ref{KS1c}) when $N=2$. 
In order that $\Delta$ becomes a usual differentiation in a proper continuum 
limit, the following condition is usually imposed:
\be
\label{KS2}
\sum_{k=0}^{N-1}a_k = 0\ ,
\ee
which is satisfied by (\ref{KS1c}). The eigenvectors for $\Delta$ are given by
\be
\label{KS3}
\phi_n^M={1 \over \sqrt{N}}\e^{i{2\pi nM \over N}}\ ,\quad M=0,1,\cdots,N-1
\ee
and their corresponding eigenvalues, by
\be
\label{KS4}
\Delta \phi_n^M = im^M \phi_n^M\ ,\quad 
im^M=\sum_{n=0}^{N-1}a_n \e^{i{2\pi nM \over N}} \ .
\ee
Here $m^0=0$. Assuming (\ref{KS1c}), one obtains
\be
\label{KS4b}
m^M=-m^{N-M}\ .
\ee
Note that $\phi_n^M$ satisfies the following properties, which may be identified 
with the conditions of normalization and completeness, respectively:
\be
\label{KS5}
\sum_{n=0}^{N-1}\phi_n^{M*} \phi_n^{M'}=\delta^{MM'}\ ,\quad 
\sum_{M=0}^{N-1}\phi_n^{M*} \phi_{n'}^{M}=\delta_{nn'}\ .
\ee
$a_n$ can be solved with respect to $m^M$ by
\be
\label{KS6}
a_n = {i \over \sqrt{N}}\sum_{M=0}^{N-1} m^M \phi_n^{M*}
= {i \over N}\sum_{M=0}^{N-1} m^M \e^{-i{2\pi nM \over N}} \ .
\ee
Then by choosing $a_n$ properly, one may obtain arbitrary spectrum of $m^M$ with $m^0=0$. 
We should note if $a_n$ is real 
\be
\label{KS6aA}
m^{N-M}=-\left(m^M\right)^*\ .
\ee
By combining (\ref{KS6aA}) with (\ref{KS4b}), $m^M$ should be real. 
For instance, if $N=3$,  $m^2=-m^1$ and 
\be
\label{KS6c}
a_0=0\ , \quad a_1 = -a_2={m_1 \over \sqrt{3}}\ .
\ee
For $N=4$, we have $m^3=-m^1$ and find $m^2=0$. 
Then 
\be
\label{KS6d}
a_0 = a_2=0\ ,\quad
a_1 = -a_3= {m_1 \over 2}\ .
\ee

The next step is to deconstruct fifth dimension. In the actions
(\ref{dSG2}-\ref{dSG4}), first replace $x^5$-dependence with $n$-dependence 
($n=0,1,\cdots,N-1$), and after that replace the derivative with respect 
to $x^5$ by the difference operator $\Delta$ (\ref{KS1}). 
It is assumed (\ref{KS1c}). First for the Lagrangian density
(\ref{dSG2}), one gets
\bea
\label{KS9}
{\cal L}_E&=& \sum_{n=0}^{N-1}\left[ -{1 \over 2}\partial_\lambda h_{n\mu\nu}
\partial^\lambda h_n^{\mu\nu}
+ \partial_\lambda h^\lambda_{n\ \mu}\partial_\nu h_n^{\mu\nu} 
 - \partial_\mu h_{n\mu\nu}\partial_\nu h_n 
+ {1 \over 2}\partial_\lambda h_n\partial^\lambda h_n \right. \nn
&& + {1 \over 2}\left(\Delta h_{n\mu\nu}\Delta h_n^{\mu\nu} - \left(\Delta h_n\right)^2
\right) - 2 \left(-\Delta B_n^\mu + \partial^\mu \varphi_n\right)
\left(\partial^\nu h_{n\mu\nu} - \partial_\mu h_n\right) \nn
&& \left. + {1 \over 2}\left(\partial_\mu B_{n\nu} - \partial_\nu B_{n\mu}\right)
\left(\partial^\mu B_n^\nu - \partial^\nu B_n^\mu\right) \right]\ .
\eea
Here and in the following,  the four-dimensional index is specified by the
Greek characters, $\mu,\nu=0,1,2,3$. In (\ref{KS9}),
\be
\label{dSG12}
B_{n\mu}=h_{n\mu 5}\ , \quad \varphi_n=h_{n55}\ .
\ee
The action $S_E=\int d^4 x {\cal L}_E$ from the Lagrangian density (\ref{KS9}) is 
invariant under transformations with the local parameters $\xi_n^\nu$ and $\zeta_n$:
\bea
\label{KS10}
h_{n\mu\nu} &\to& h_{n\mu\nu} + \partial_\mu \xi_{n\nu} + \partial_\nu \xi_{n\mu}\ ,\nn
B_{n\mu} &\to& B_{n\mu} + \Delta \xi_{n\mu} - \partial_\mu \zeta_n\ ,\nn
\varphi_n &\to& \varphi_n - \Delta \zeta_n\ ,
\eea
which comes from the general coordinate transformation in (\ref{dSG6}). 
For the Lagrangian of the Rarita-Schwinger field (\ref{dSG3}), we have
\bea
\label{dSG13}
{\cal L}_{RS} &=& \sum_{n=0}^{N-1}\left\{i\bar \psi_{n\mu} \gamma^{\mu\nu\rho}
\partial_\nu \psi_{n\rho}
+i\bar \psi_{n5} \gamma^{5\nu\rho}\partial_\nu \psi_{n\rho} \right. \\
&& \left. + i\bar \psi_{n\mu} \gamma^{\mu 5\rho} \Delta \psi_{n\rho} 
+ i\bar \psi_{n\mu} \gamma^{\mu\nu 5} \partial_\nu \psi_{n5}\right\}\ .
\eea
The action $S_{RS}=\int d^4 x {\cal L}_{RS}$ is invariant under the transformation
with $N$-local fermionic parameters $\eta_n$ $\left(n=0,1,\cdots N-1\right)$:
\be
\label{dSG14}
\delta \psi_{n\mu} = \partial_\mu \eta_n\ ,\quad \delta \psi_{n5} 
= \Delta \eta_n\ ,
\ee
which correspond to the local supersymmetry transformation in (\ref{dSG7}).
With the redefinition
\be
\label{dSG15}
\psi_{n\mu}'\equiv \psi_{n\mu} + {1 \over 2}\gamma_\mu \gamma_5 \psi_{n5}\ ,\quad 
\psi_n'\equiv \sqrt{3 \over 2} \psi_{n5}\ ,
\ee
the Lagrangian density (\ref{dSG13}) can be rewritten as
\bea
\label{dSG16}
{\cal L}_{RS} &=& \sum_{n=0}^{N-1}\left\{i\bar \psi_{n\mu}' \gamma^{\mu\nu\rho}
\partial_\nu \psi_{n\rho}' 
+i\bar \psi_{n}' \gamma^{\nu}\partial_\nu \psi_{n}'
 - i\bar \psi_{n\mu}' \gamma^5\gamma^{\mu\rho} \Delta \psi_{n\rho}' \right. \\
&& \left. -i\sqrt{3 \over 2}\bar \psi_{n}' \gamma^\rho \Delta \psi_{n\rho}' 
 - i\sqrt{3 \over 2} \bar \psi_{n\mu}' \gamma^\mu \Delta \psi_n'
 -2i\bar \psi_n' \Delta \psi_n'\right\}\ .
\eea
In terms of the redefined fields, the transformation  (\ref{dSG14}) is rewritten as
\be
\label{dSG17}
\delta \psi_{n\mu}' = \partial_\mu \eta_n + {1 \over 2}\gamma_\mu \gamma_5 \Delta \eta_n\ ,
\quad \delta \psi_n' = \sqrt{3 \over 2}\Delta \eta_n\ .
\ee
By using the transformation (\ref{dSG17}), the following gauge condition
can be chosen
\be
\label{dSG18}
\gamma^\mu \psi_{n\mu}'=0\ .
\ee
Then the action reduces to the sum of the action of the (4-dimensional) Rarita-Schwinger 
field and that of the Dirac spinor fields:
\be
\label{dSG19}
{\cal L}_{RS} = \sum_{n=0}^{N-1}\left\{i\bar \psi_{n\mu}' \gamma^{\mu\nu\rho}
\partial_\nu \psi_{n\rho}' +i\bar \psi_{n}' \gamma^{\nu}\partial_\nu \psi_{n}' 
+ i\bar \psi_{n\mu}' \gamma^5\gamma^{\mu\rho} \Delta \psi_{n\rho}' 
 -2i\bar \psi_n' \Delta \psi_n'\right\}\ .
\ee
Finally from the Lagrangian density (\ref{dSG4}) of the vector field, we obtain
\be
\label{dSG20} 
{\cal L}_U = - {1 \over 4}F_n^{\mu\nu}F_{n\mu\nu} - {1 \over 2}
\left(\partial^\mu \rho_n - \Delta A_n^\mu\right)
\left(\partial_\mu \rho_n - \Delta A_{n\mu}\right)\ ,
\ee
This is invariant under the transformation, which corresponds to $U(1)$
gauge transformation (\ref{dSG8}):
\be
\label{dSG21}
\delta A_{n\mu} = \partial_\mu \sigma_n\ ,\quad \delta \rho_n = \Delta \sigma_n\ .
\ee
In (\ref{dSG20}), 
\be
\label{dSG22}
\rho_n \equiv A_{n5}\ .
\ee
In terms of the redefined fields in (\ref{dSG12}), (\ref{dSG15}), and (\ref{dSG22}), the 
supersymmetry transformation (\ref{dSG9}) can be rewritten as
\bea
\label{dSG23}
\delta h_{n\mu\nu}&=& {1 \over 2}\left( \bar\zeta \gamma_\mu \psi_{n\nu}' 
+ \bar\zeta \gamma_\nu \psi_{n\mu}'  - \bar \psi_{n\mu}' \gamma_\nu \zeta 
 - \bar \psi_{n\nu}' \gamma_\mu \zeta\right)\nn
&& + {1 \over \sqrt{6}}\eta_{\mu\nu}\left(-\bar\zeta\gamma_5\psi_n' 
+ \bar\psi_n'\gamma_5 \zeta\right)\ ,\nn 
\delta B_{n\mu}&=& {1 \over \sqrt{6} }\left(\bar\zeta\gamma_\mu\psi_n' 
 - \bar\psi_n'\gamma_\mu\zeta\right) + {1 \over 2}\left(\bar\zeta\gamma_5\psi_{n\mu}' 
 -\bar\psi_{n\mu}'\gamma_5 \zeta\right) + {1 \over 2\sqrt{6}}\left(\bar\zeta\gamma_\mu\psi_n' 
 - \bar\psi_n'\gamma_\mu\zeta\right)\ ,\nn
\delta\varphi_n&=& \sqrt{2 \over 3}\left(\bar\zeta\gamma_5 \psi_n' 
 - \bar\psi_n'\gamma_5 \zeta\right)\ ,\nn
\delta\psi_{n\mu}'&=& - {i \over 4}\gamma^{\rho\sigma} \omega_{n\mu\rho\sigma}\zeta 
 - {i \over 4}\gamma^5 \gamma^\sigma \left(\Delta h_{n\sigma\mu} 
 - \partial_\sigma B_{n\mu}\right) \nn
&& + F_{n\rho\sigma}\left(\gamma^{\rho\sigma}\gamma_\mu - 5 \gamma^\rho \delta_\mu^{\ \sigma} 
+ 5 \gamma^\sigma \delta_\mu^{\ \rho}\right) 
+ 2\left(\Delta A_{n\sigma} - \partial_\sigma\rho\right)\left(\gamma^5\gamma^\sigma 
\gamma_\mu - 5\gamma^5 \delta_\mu^{\ \sigma}\right) \ ,\nn
\delta\psi_n'&=&\sqrt{3 \over 2}\left\{-{i \over 8}\gamma^{\rho\sigma}\zeta\left(
\partial_\rho B_{n\sigma} - \partial_\sigma B_{n\rho}\right) 
 - {i \over 8}\gamma^5 \gamma^\sigma\zeta \left(\Delta B_{n\sigma}
 - \partial_\sigma \varphi\right) \right. \nn
&& \left. + F_{n\rho\sigma}\gamma^{\rho\sigma}\gamma_5 
+ 9\left(\Delta A_{n\sigma} - \partial_\sigma\rho\right)\gamma^\sigma \right\}\ ,\nn
\delta A_{n\mu} &=& 80\left(\bar\zeta\psi_{n\mu}' + \bar\psi_{n\mu}'\zeta\right) 
 - {80 \over \sqrt{6}}\left(\bar\zeta \gamma_\mu\gamma_5\psi_n' 
+ \bar\psi_n' \gamma_5 \gamma_\mu \zeta\right)\ , \nn
\delta\rho &=& 80\sqrt{2 \over 3}\left(\bar\zeta\psi_n' + \bar\psi'\zeta\right)\ .
\eea
Here $\omega_{n\mu\nu\rho}= {1 \over 2} \left\{\partial_\mu h_{n\rho\nu}
 - \partial_\rho h_{n\mu\nu}\right\}$ and  the gauge condition (\ref{dSG5c})
is used. This finishes the construction of linearized multisupergravity 
from discrete extra dimension. Note that such theory is free of ghosts 
(like the linearized multigravity) and most of symmetries of five-dimensional
supergravity are respected. To address the consistency (ghosts presence) 
one needs to go beyond the linear level which is quite non-trivial task.

\section{One-loop vacuum energy}

Now the on-shell degrees of the freedom may be counted. As clear from
(\ref{KS4}), $\Delta$ gives the mass. First one considers the massless particles. 
In the Lagrangian density (\ref{KS9}), the massless particles are the graviton 
$h_{n\mu\nu}$, vector field $B_{n\mu}$, and scalar field $\varphi_n$. The on-shell 
degrees of the freedom are 2, 2, and 1, respectively.  
In the Lagrangian density (\ref{dSG16}) or (\ref{dSG19}), the massless particles 
are the complex Rarita-Schwinger field $\psi_{n\mu}'$ and the Dirac fermion $\psi_n'$, 
whose on-shell degrees of the freedom are $2\times 2$ and $2\times 2$. 
Finally in the Lagrangian density (\ref{dSG20}), the vector field $A_{n\mu}$ 
and the scalar field $\rho_n$ are massless and their physical degrees of the freedom 
are 2 and 1, respectively. Then in the massless sector, the total number of on-shell 
degrees of the freedom is 8 in the both of the bosonic and the fermionic sector. 
In the massive sector, several fields can be eliminated by the local symmetry 
transformation. In (\ref{KS9}) by using the linearized general coordinate 
transformation (\ref{KS10}),  vector field $B_{n\mu}$ and scalar field 
$\varphi_n$ can be eliminated. The remaining field is massive graviton
$h_{n\mu\nu}$, whose on-shell number of degrees of the freedom is 5 as $h_{n\mu\nu}$ 
has spin 2. For the massive particles in (\ref{dSG16}), we may choose the gauge 
condition $\psi_n'=0$, instead of (\ref{dSG18}), by using the local supersymmetry 
transformation (\ref{dSG17}). Then the remaining massive complex spin $3/2$ particle 
$\psi_{n\mu}'$ has the on-shell degrees of freedom $4\times 2$.
In (\ref{dSG20}), one may eliminate $\rho_n$ in the massive sector by
using the gauge transformation (\ref{dSG21}) and the remaining massive vector 
(spin 1) particle has 3 on-shell degrees of the freedom. Then even in the massive 
sector with common mass, the total number of on-shell degrees of the freedom 
is 8 in the both of the bosonic and the fermionic sector. As the on-shell degrees 
of the freedom in the bosonic sector coincide with those in the fermionic sector 
and now we are considering the flat background, the one-loop vacuum energy coming 
from the bosonic sector cancells with that from the fermionic sector. 

For the fermionic sector, one may impose anti-periodic boundary condition
for the discretized fifth dimension
\be
\label{dSG24}
\psi_{n+N\,\mu}'=-\psi_{n\mu}'\ ,\quad \psi_{n+N}'=-\psi_n'\ .
\ee
Then instead of (\ref{KS3}), the eigenvectors for $\Delta$ are given by
\be
\label{KS3A}
\phi_{An}^M={1 \over \sqrt{N}}\e^{i{2\pi n\left(M+{1 \over 2}\right) \over N}}\ ,\quad 
M=0,1,\cdots, N-1
\ee
and their corresponding eigenvalues, by
\be
\label{KS4A}
\Delta \phi_{An}^M = im_A^M \phi_{An}^M\ ,\quad 
im_A^M=\sum_{n=0}^{N-1}a_n \e^{i{2\pi n\left(M + {1 \over 2}\right) \over N}} \ .
\ee
As in (\ref{KS5}), $\phi_{An}^M$ satisfies the conditions of normalization and completeness:
\be
\label{KS5A}
\sum_{n=0}^{N-1}\phi_{An}^{M*} \phi_{An}^{M'}=\delta^{MM'}\ ,\quad 
\sum_{M=0}^{N-1}\phi_{An}^{M*} \phi_{An'}^{M}=\delta_{nn'}\ .
\ee
If the fermionic particles obey the anti-periodic boundary condition, the (global) 
supersymmetry is explicitly broken, what becomes manifest in the mass
spectrum. It is interesting that for continious orbifold fifth dimension
the antiperiodic boundary conditions maybe interpreted as a discrete 
Wilson line breaking of supersymmetry (see, for instance \cite{mariano}).

In general, the one-loop vacuum energy of the real scalar with mass $m$ can be 
evaluated by the $\zeta$-function regularization\cite{CENOZ}\footnote{The 
related discussion of matter Casimir effect in deconstruction maybe found in \cite{BLS}}:
\be
\label{dSG25}
V^b_{\rm eff}=V^b_R(\mu) + {m^4 \over 64\pi^2}\left(\ln{m^2 \over \mu^2} - {3 \over 2}\right)\ .
\ee
Here $\mu$ is introduced for the renormalization. $V_R(\mu)$ is determined by the condition 
that $V_{\rm eff}$ should not depend on the arbitrary parameter $\mu$:
\be
\label{dSG26}
\mu{dV_{\rm eff} (\mu) \over d\mu}=0\ .
\ee
As we are now considering the flat background, the contribution to the one-loop vacuum 
energy from each of the bosonic degrees of freedom is given by (\ref{dSG25}). On the other 
hand, the contribution $V^f_{\rm eff}$ from each of the fermionic degrees of the freedom 
is different from that from bosonic ones by sign:
\be
\label{dSG27}
V^f_{\rm eff}=-V^b_{\rm eff}\ .
\ee
If the fermionic particles satisfy the periodic boundary condition, 
the mass spectrum and the degrees of the freedom of the fermionic particle are 
identical with those of the bosonic particles. Then the one-loop vacuum energy 
vanishes, which is also a signal of the supersymmetry. On the other hand,
if the boundary condition for fermionic fields is anti-periodic, the 
vacuum energy does not vanish in general. 
As an example, we consider $N=3$ case in (\ref{KS6c}). 
Then for the bosonic sector, the mass $m_b$ can be
\be
\label{dSG28}
m_b^2=\left(0,m_1^2, m_1^2\right)\ .
\ee
For the fermionic sector, by using (\ref{KS4A}), we find the mass $m_f$ for 
the fermionic sector is given by
\be
\label{dSG29}
\left|m_f\right|^2=\left( {m_1^2 \over 3}, {4m_1^2 \over 3}, {4m_1^2 \over 3}\right)\ .
\ee
Due to the anti-periodic boundary condition, there is no massless fermionic particle. 
The total effective potential $V_{\rm eff}$ is given by
\bea
\label{dSG30}
V_{\rm eff}&=& V_R(\mu) + {16m_1^4 \over 64\pi^2}\left(\ln{m_1^2 \over \mu^2} - {3 \over 2}
\right) \nn
&& - 8\left\{{2m_1^4 \over 9\cdot 64\pi^2}\left(\ln{m_1^2 \over 3\mu^2} - {3 \over 2}
\right) + {16m_1^4 \over 9\cdot 64\pi^2}\left(\ln{4m_1^2 \over 4\mu^2} - {3 \over 2}
\right)\right\} \nn
&=& V_R(\mu) - {m_1^4 \over 36\pi^2}\left(16\ln 2 - \ln 3\right)\ .
\eea
 Eq.(\ref{dSG26}) gives $V_R(\mu)=0$ and 
\be
\label{dSG31}
V_{\rm eff}= - {m_1^4 \over 36\pi^2}\left(16\ln 2 - \ln 3\right)<0\ .
\ee
In the lattice field theory, $a_1$ in (\ref{KS6c}) is related with the lattice spacing 
$a$ by $a_1={1 \over 2a}$. As a result, $m_1={\sqrt{3} \over 2a}$. If we regard the 
sites on the lattice with the branes in the extra dimension, the lattice spacing
$a$ may correspond to the distance between the branes. 

In getting (\ref{dSG31}), we have assumed that the bosonic particles obey the periodic 
boundary condition as we consider the lattice on the circle $S^1$. 
However, taking the lattice on the orbifold obtained by dividing $S^1$ with discrete 
subgroup $Z_2$ (with the coordinate on the circle as $\theta$, $0\leq \theta < 2\pi$, 
$S^1/Z_2$ can be obtained by identifying $\theta$ with $\theta+\pi$), the bosonic 
particle can obey the anti-periodic boundary condition. Thus, for the anti-periodic 
boundary condition for the bosonic particles and the periodic one for the fermionic 
particles, the sign of the vacuum energy (\ref{dSG31}) is reversed and one obtains 
the effective positive cosmological constant (dark energy) \cite{CENOZ}, which may 
explain the current universe acceleration. 

\section{Discussion}

The resulting theory contains multigravitons and gravitinos. In this sense, we may 
call it multisupergravity. However, this is linearized model which does not
include the interaction. In order to obtain the complete 
supergravity theory with the interaction, we may start from the local supersymmetry 
transformation which is given by combining the linearized local supersymmetry 
transformation (\ref{dSG17}) with the transformation (\ref{dSG24}) after replacing the 
constant spinor parameter $\zeta$ with the local parameter $\eta_n$. The variation 
of the total Lagrangian density ${\cal L}={\cal L}_E + {\cal L}_{RS} + {\cal L}_U$ is not 
invariant but proportional to $\partial_\mu \eta_n$ or $\partial_\mu \bar\eta_n$ 
up to the total derivative. One may add the term obtained by replacing 
$\partial_\mu \eta_n$ ($\partial_\mu \bar\eta_n$) in the variation with 
$-\psi_{n\mu}'$ $-\left(\bar\psi_{n\mu}'\right)$ to the action as a counterterm. 
The counterterm includes interaction in general. The modified Lagrangian
is not still invariant in general but its variation is proportional to 
$\partial_\mu \eta_n$ or $\partial_\mu \bar\eta_n$ up to the total derivative again.
More counterterms may be added. This is nothing but the Noether method. 
If the procedure ends up by the finite number of the iterations, the complete 
supergravity model results. In general, however, the procedure does not 
end up with the finite iteration. Instead, as in \cite{AGS}, we may start 
from the $N$-copies of some kind of four-dimensional supergravity (with
interaction) and introduce matter multiplet linking the copies. As a result, 
one may construct an interacting theory with massive graviton and gravitino(s). 
Linearizing such a theory, one gets the model of the sort discussed in this paper. 
It could be that complete discretized supergravity (for which the theory under 
consideration is just the first, preliminary step) may be useful to solve 
the consistency problem of multigravity (see, however, example of ghost free 
bi-gravity in\cite{padilla}). Unfortunately, its construction is 
highly non-trivial problem. Note, however, that after first submission of this work to hep-th the very interesting approach which admits the construction of
 complete multisupergravity in terms of superfields has been developed \cite{GSS}.

Although the five-dimensional linearized discretized supergravity is
considered, the generalization to higher dimensions may be done.
For example, if we start from the eleven-dimensional supergravity \cite{BST} 
and replace the eleventh direction with the discrete lattice, 
we may obtain the ten-dimensional supergravity theories with multigravitons and 
multigravitinos. This may open new interesting connection between M-theory 
and ten-dimensional supergravities (strings). 

\section*{Acknowledgments}

We are very grateful to Mariano Quiros and Paul Townsend for stimulating discussions.
This investigation has been supported in part by the Ministry of
Education, Science, Sports and Culture of Japan under the grant n.13135208
(S.N.), by RFBR grant 03-01-00105 (S.D.O.) and by LRSS grant 1252.2003.2 (S.D.O.). 

\appendix

\section{$\gamma$-matrices}

Here, the conventions and the basic formulae for the five-dimensional 
$\gamma$-matrices are summarized.

The definition of the $\gamma$-matrices is
\be
\label{g1}
\left\{ \gamma^\mu, \gamma^\nu \right\}=2\eta^{\mu\nu}\ .
\ee
$\gamma^0$ is anti-hermitian,  $\gamma^i$ $\left(i=1,2,3,4\right)$'s are 
hermitian and $\gamma_0 \left(\gamma^\mu\right)^\dagger \gamma_0 = \gamma^\mu$. 
For the spinor field $\psi$, we define $\bar\psi \equiv i \psi^\dagger \gamma_0$. 
As a result, $\left(\bar\psi_1 \psi_2\right)^\dagger
= \bar\psi_2 \psi_1$, $\left(\bar\psi_1 \gamma^\mu \psi_2\right)^\dagger
=-\bar\psi_2 \gamma^\mu \psi_1$, etc. 
We also define $\gamma^{\mu\nu}$ and $\gamma^{\mu\nu\rho}$ by
\bea
\label{g2}
\gamma^{\mu\nu}&\equiv&{1 \over 2}\left( \gamma^\mu \gamma^\nu 
 - \gamma^\nu \gamma^\mu\right)\ ,\nn
\gamma^{\mu\nu\rho}&\equiv&{1 \over 6}\left( \gamma^\mu \gamma^\nu\gamma^\rho
+ \gamma^\nu \gamma^\rho\gamma^\mu + \gamma^\rho \gamma^\mu\gamma^\nu 
 - \gamma^\nu \gamma^\mu\gamma^\rho  - \gamma^\mu \gamma^\rho\gamma^\nu 
 - \gamma^\rho \gamma^\nu\gamma^\mu \right)\ .
\eea
where
\be
\label{g3}
\gamma^{\mu\nu\rho}=-i \epsilon^{\mu\nu\rho\sigma\tau}\gamma_{\sigma\tau}\ .
\ee
Here $\epsilon^{\mu\nu\rho\sigma\tau}$ is rank 5 anti-symmetric tensor and 
$\epsilon^{01234}=1$. The following relations are necesssary to show the 
invariance under the global supersymmetry transformation (\ref{dSG9}):
\bea
\label{g4}
\gamma^{\eta\zeta}\gamma^\sigma &=& \gamma^{\eta\zeta\sigma} + \gamma^\eta \eta^{\sigma\zeta} 
- \gamma^\zeta \eta^{\sigma\eta}\ ,\nn
\gamma^{\mu\nu\rho} \gamma^{\sigma\tau}
&=& -4i\epsilon^{\mu\nu\rho\sigma\tau} - 4 \left(
 - \eta^{\rho\sigma} \eta^{\mu\tau} \gamma^\nu + \eta^{\mu\sigma} \eta^{\rho\tau}\gamma^\nu 
 - \eta^{\mu\sigma} \eta^{\nu\tau} \gamma^\rho + \eta^{\nu\sigma} \eta^{\mu\tau} \gamma^\rho 
\right.\nn && \left.
 - \eta^{\nu\sigma} \eta^{\rho\tau} \gamma^\mu 
+ \eta^{\rho\sigma}\eta^{\nu\tau} \gamma^\mu \right) \nn
&& -2 \left(\eta^{\mu\sigma}\gamma^{\nu\rho\tau} + \eta^{\nu\sigma}\gamma^{\rho\mu\tau} 
+ \eta^{\rho\sigma}\gamma^{\mu\nu\tau}\right) 
+ 2\left(\eta^{\mu\tau}\gamma^{\nu\rho\sigma} + \eta^{\nu\tau}\gamma^{\rho\mu\sigma} 
+ \eta^{\rho\tau}\gamma^{\mu\nu\sigma}\right) \nn
&& + 2i\epsilon^{\mu\nu\rho\eta\sigma}\gamma_\eta^{\ \tau} 
 -2i \epsilon^{\mu\nu\rho\eta\tau}\gamma_\eta^{\ \sigma} \ , \nn
\gamma^{\mu\nu\rho} \gamma^\sigma
&=& -4\left(\eta^{\mu\sigma}\gamma^{\nu\rho} + \eta^{\rho\sigma}\gamma^{\mu\nu}
+ \eta^{\sigma\nu}\gamma^{\rho\mu}\right)
+ 2i\epsilon^{\mu\nu\rho\sigma\tau}\gamma_\tau\ .
\eea

\end{document}